\documentclass[prb,aps,twocolumn]{revtex4}
\usepackage{amsmath}
\usepackage{graphicx}
\begin{document}

\title{Spontaneous vortex state and $\varphi$-junction in a superconducting bijunction with a localized spin}

\date{\today} 

\author{Denis Feinberg$^{1,2}$, C. A. Balseiro$^{3,4}$} 
\affiliation{$^1$ Centre National de la Recherche Scientifique, Institut NEEL, F-38042 Grenoble Cedex 9, France}
\affiliation{$^2$ Universit\'e Grenoble-Alpes, Institut NEEL, F-38042 Grenoble Cedex 9, France}
\affiliation{$^3$ Centro At\'omico Bariloche and Instituto Balseiro, Comisi\'on Nacional de Energia At\'omica, 8400 Bariloche, Argentina}
\affiliation{$^4$ Consejo Nacional de Investigaciones Cient\'ificas y T\'ecnicas (CONICET), Argentina}

\begin{abstract}
A  Josephson bijunction made of three superconductors connected by a quantum dot is considered in the regime where the dot carries a magnetic moment. In the range of parameters where such a dot, if inserted in a two-terminal Josephson junction, creates a $\pi$-shift of the phase, the bijunction forming a triangular unit is frustrated. This frustration is studied both within a phenomenological and a microscopic model. Frustration stabilizes a phase vortex centered on the dot, with two degenerate states carrying opposite vorticities, independently of the direction of the magnetic moment. Embedding the bijunction in a superconducting loop allows to create a tunable "$\varphi$"-junction whose equilibrium phase can take any value. For large enough inductance, it generates noninteger spontaneous flux. Multi-loop configurations are also studied.  
\end{abstract}

\maketitle

\section{Introduction}
Junctions with confined electrons, like atomic, molecular or quantum-dot (QD) junctions, are among the most studied nanoscopic devices \cite{Andergassen}. In these structures electron-electron correlations within the junction together with the electronic properties of the contacts lead to a number of different phenomena concerning fundamental aspects of quantum charge and spin transport.  
Modern technologies allowed building junctions that are close to the ultimate limit of miniaturization with normal, superconducting \cite{Yu, Morpurgo, Buitelaar, Doh, Jorgensen, Xiang, Jarillo} or ferromagnetic \cite{Ferro1, Ferro2} leads creating new opportunities for novel nanodevices with predefined functional properties.  One-electron transistors, spin valves or superconducting spin qubits are some examples of such devices.
When a molecule or a QD bridges the gap between two metallic leads, the Coulomb energy tends to quantize the charge confined in the junctions, i.e. on the molecule or the dot, leading to the possibility of fabricating a junction with a confined spin. In the case of superconducting leads, the characteristic of the so-obtained Josephson junction depends on the total spin in the junctions. 
Superconducting circuits with QD (S-QD-S junctions) have been extensively studied during the last decade  \cite{Levy_Yeyati_Review,De_Franceschi_Review}. These junctions are fabricated by contacting superconducting leads to a normal nanostructure, typically a single walled carbon nanotube or a semiconducting nanowire. The structure may include gate electrodes that can be used to control the number of electrons in the dot. 
S-QD-S junctions having a localized spin in the dot may have a global minimum of the free energy for a $\pi$-difference between the phases of the two superconducting contacts . The current-phase characteristic of these junctions is described by a Josephson equation with a {\it negative} critical current \cite{pi_junction1, pi_junction2}.  These junctions, referred to as $\pi$-junctions, in contrast with standard "$0$-junctions", have interesting properties and potential applications in superconducting electronics, including phase or flux qubits \cite{Makhlin_Review}.
A superconducting ring containing a $\pi$-junction could generate a spontaneous current with (nearly) half a superconducting quantum flux threading the ring, a very convenient situation for experimental detection. Notice that this current structure is stable only if the ring self-inductance exceeds a critical value\cite{Sigrist_Rice,Jagla}. 

Here we present results for a $\pi$-bijunction consisting of a QD connected to three superconducting leads, see Fig. \ref{fig:Triterminal_dot}. Graphene dots (GQD) are good candidates to build such a device\cite{graph1, graph2}. In fact graphene offers new opportunities for superconducting electronics as a new class of material that can be tailored and contacted to normal or superconducting leads. With gate electrodes controlling the number of electrons confined in the GQD, a non-zero spin can be localized at the dot which tends to generate a $\pi$-shift between each of the pairs of superconductors. This situation, in a way similar to Heisenberg magnets, generates frustration\cite{Sigrist_Rice}.The frustration can be resolved by canting the phases of the three superconductors, in a way that depends on the asymmetry of the device. This asymmetry, due to different couplings between the dot and the contacts, can be controlled by gates. Frustration leads to canting only for moderate asymmetries. In the canted phase, the equilibrium phase difference between two given superconductors can be controlled at will to a value $\varphi_m$, between $0$ and $\pi$. Such a tunable $\varphi$-junction can be probed by various geometries incorporating one or several superconducting loops. An important feature of the phase canted state is that it contains a spontaneous vorticity. Due to time-reversal symmetry, two equivalent solutions with opposite vorticities are found, corresponding to phases $\varphi_m$ and $-\varphi_m$. This symmetry can be broken with the help of a single loop and an applied magnetic field. In addition, the structure of the energy-phase profile of the bijunction makes the barrier between the two degenerate minima tunable, either through the bijunction parameters, or using the external flux. This might be an useful property for building a superconducting qubit.

The paper is organized as follows. In Section 2 a phenomenological phase model is first considered, with a stability analysis of the canted (frustrated) solution against the non-frustrated solution. Then a microscopic model for a dot with a single level allows a nonperturbative solution, which essentially confirms the existence of a canted phase below a critical asymmetry of the dot-contact couplings. Section 3 considers a single loop set-up, with an applied orbital magnetic field, then two-loop or three-loop set-ups.

\section{bijunction at equilibrium}

\subsection{A phenomenological model}

\begin{figure}[htb]
\includegraphics[width=1.0\columnwidth]{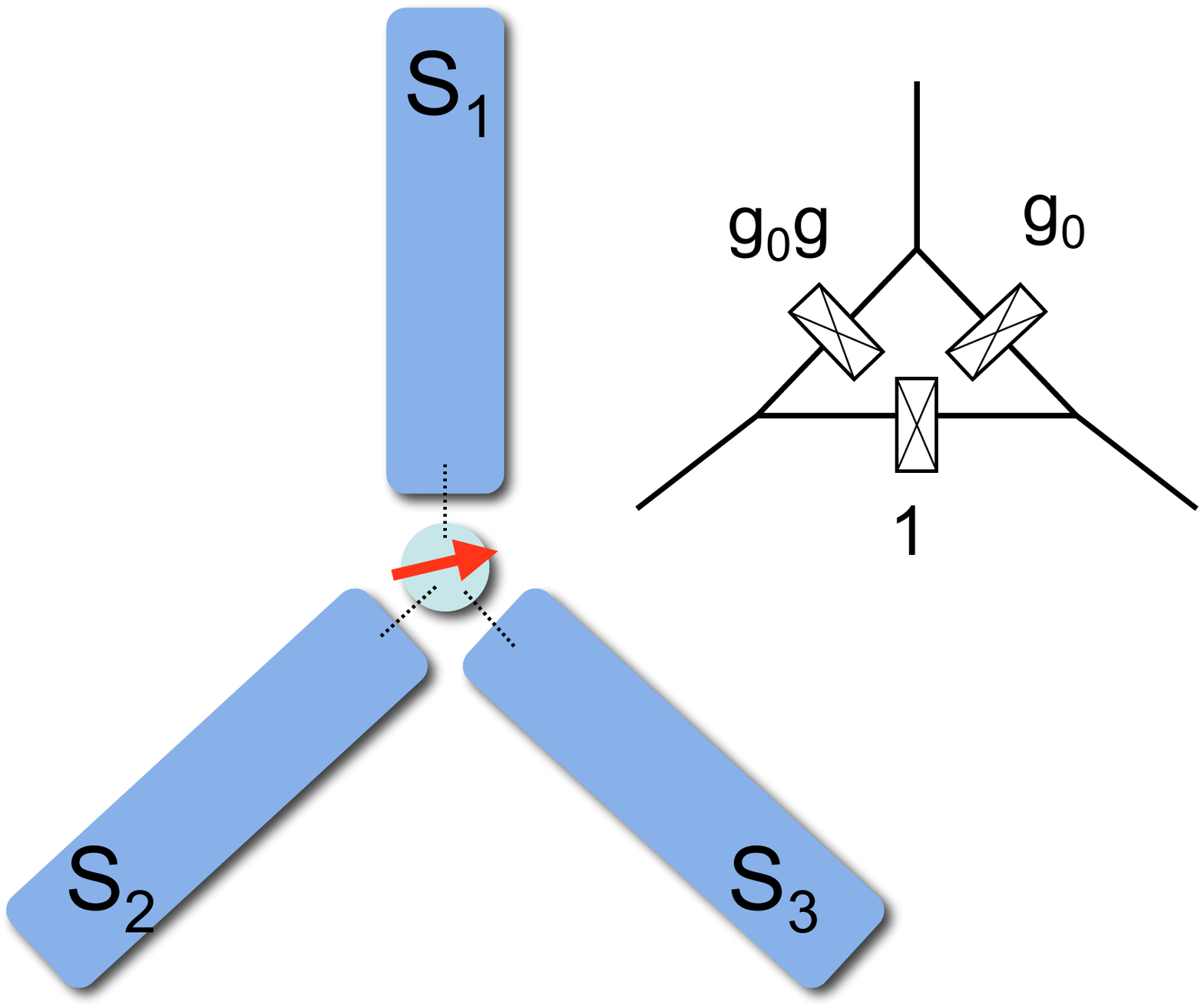}
\caption{(Color online) Bijunction made of three superconductors and one quantum dot carrying a spin $S=1/2$. Inset show the equivalent triangular model, valid in the perturbative regime only. The asymmetry ratios between the junctions critical currents are indicated.
\label{fig:Triterminal_dot}
}
\end{figure}

The quantum dot connects all three superconductors (Fig. \ref{fig:Triterminal_dot}). Each pair $(i,j)$ of the three superconductors ($i=1,2,3$) forms a Josephson junction. As a first approximation, one can write the total energy of the bijunction as the sum of those of separated junctions. Such an expression could be obtained in perturbation theory from a microscopic Hamiltonian, at fourth order in the tunnelling element between the superconductor states and the dot states. The bijunction is then equivalent to a triangular array of separated junctions.  We assume that the presence of a $1/2$ spin on the dot creates $\pi$-junctions, and  that this holds for all of them. The nonperturbative calculation presented in the next subsection shows that it is essentially the case, unless the couplings to the dot are very asymmetric. Denoting the superconducting phases as $\varphi_i$ ($i=1-3$), the bijunction energy thus reads :

\begin{equation}
\label{eq:energy_BJ_red}
E_{BJ}=E_{0}[g_0g\cos\varphi_{12}+g_0\cos\varphi_{13}+\cos\varphi_{23}],
\end{equation}
with $\varphi_{ij}=\varphi_i-\varphi_j$, $E_0>0$ and where $g_0\geq0$, $g\geq0$ are parameters quantifying the bijunction asymmetry (Fig. \ref{fig:Triterminal_dot}). 

Let us  look for the equilibrium state. Setting to zero the partial derivatives of $E_{BJ}$ with respect to the $\varphi_i$'s is equivalent to imposing zero current $J_i$ in each lead $S_i$. One obtains from $J_1=0$ :

\begin{equation}
\label{zerocurrent}
g\sin\varphi_{12}+\sin\varphi_{13}=0,
\end{equation}
which, together with similar equations expressing that $J_2$ or $J_3=0$, yields :

\begin{equation}
\label{canting}
\cos\varphi_{23}=\frac{(g_0g)^2-1-g^2}{2g}.
\end{equation}

\begin{figure}[htb]
\includegraphics[width=1.0\columnwidth]{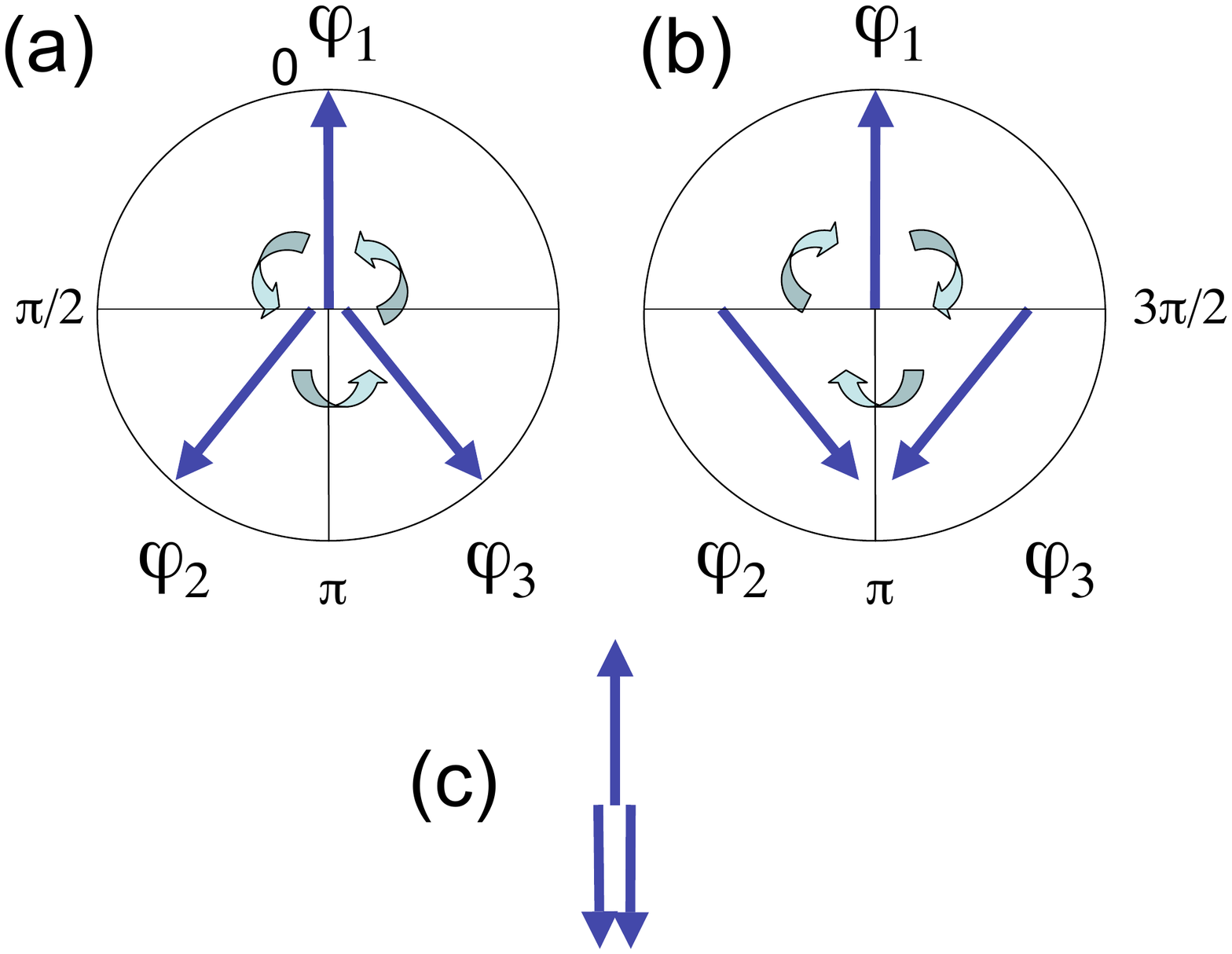}
\caption{(Color online) a-b)Two symmetric canted solutions in the frustrated situation, corresponding to a spontaneous phase vortex. The straight arrows represent the superconducting phase. c) non frustrated solution with two $\pi$-junctions ($S_1-S_2$ and $S_1-S_3$) and a $0$-junction ($S_2-S_3$).
\label{fig:phase_vortices}
}
\end{figure}

Such a nontrivial solution thus exists only if $|1-\frac{1}{g}|\leq g_0 \leq (1+\frac{1}{g})$. This is a canted (e.g. frustrated) phase solution (Fig. \ref{fig:phase_vortices}a,b), with two degenerate states obtained from each other by changing $\varphi_{2,3}$ into $-\varphi_{2,3}$. In such states, the current across any junction $S_i-S_j$ is nonzero. Yet, the total current in each lead is zero. Those two degenerate solutions therefore feature a phase vortex, with two opposite vorticities. While in a real triangular network this vorticity is associated to a circulating current, with a zero-dimensional quantum dot, it is difficult to define a path with a nonzero current circulating around the dot. Nevertheless, we show in the last Section that a true vortex can be induced on an adjacent loop.

In the opposite case $|(g_0g)^2-1-g^2|\geq 2g$, the energy minimum is obtained for $\varphi_2=0$ or $\pi$, $\varphi_3=0$ or $\pi$, replacing Eq. (\ref{canting}). This results in two of the three junctions being $\pi$-junctions and the other one a $0$-junction (Fig. \ref{fig:phase_vortices}c). 

Later on we consider situations where the lead $1$ is disconnected (e.g. the phase $\varphi_1$ is floating), while leads $2$ and $3$ are connected to an external circuit. Then the convenient phase variable is $\varphi=\varphi_2-\varphi_3$. One can use gauge invariance and choose $\varphi_2=\frac{\varphi}{2}$, $\varphi_3=-\frac{\varphi}{2}$. Then Equation (\ref{zerocurrent}) yields 

\begin{equation}
\tan\varphi_1=G\tan(\frac{\varphi}{2}).
\end{equation}
with $G=\frac{g-1}{g+1}$. The total energy reads :

\begin{equation}
\label{eq:energy_TJ}
E_{TJ}=E_{0}\Big [\cos\varphi - g_0(g+1)\frac{|\cos(\frac{\varphi}{2})|}{\sqrt{1+G^2\tan^2(\frac{\varphi}{2})}}\Big ].
\end{equation}

\begin{figure}[htb]
\includegraphics[width=.8\columnwidth]{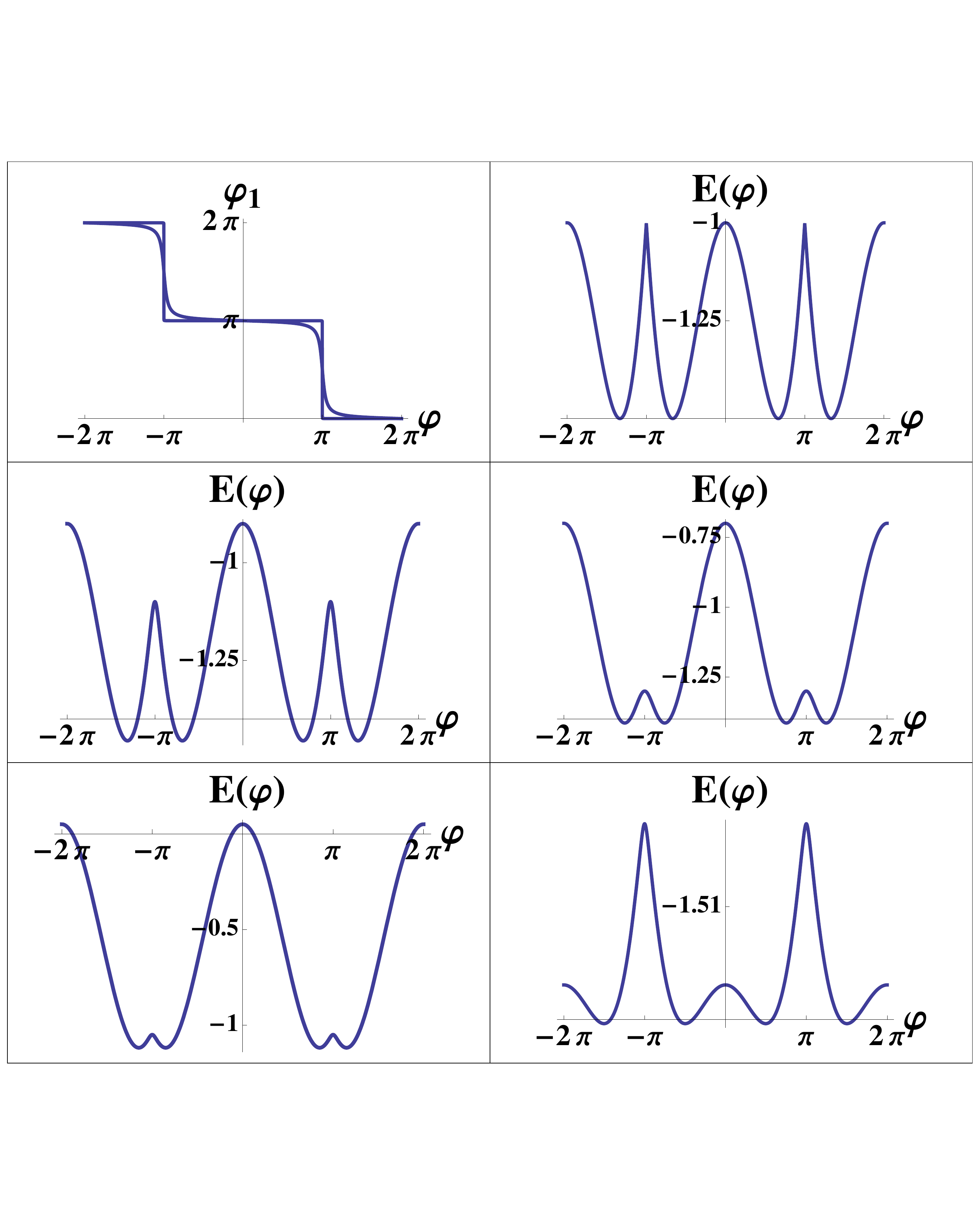}
\caption{(Color online). (Top left) Variation of the "floating" phase $\varphi_1$ with the phase $\varphi$ across $S_2-S_3$, for $g=1$ (staircase) and $0.9$. Total energy of the bijunction inserted in a single loop, as a function of the phase $\varphi_{23}=\varphi$ for (top right) $g_0=1$, $g=1$; (middle left) $g_0=1$, $g=0.9$; (middle right) $g_0=1$, $g=0.7$; (bottom left) $g_0=0.5$, $g=0.9$; (bottom right) $g_0=1.5$, $g=0.9$. Despite the $2\pi$-periodicity, the plot between $\varphi=\pm  2\pi$ shows the different barriers depending on the couple of degenerate phase vortex states.
\label{fig:phenomenol_sans_flux}
}
\end{figure}

The variation of $\varphi_1$ with $\varphi$ (see Fig. \ref{fig:phenomenol_sans_flux}) shows that for a partially symmetric bijunction ($g=1$),
 $\varphi_1$ jumps by $\pi$ each time $\varphi$ is an odd multiple of $\pi$. The energy profile of the bijunction is pictured on Fig.
 \ref{fig:phenomenol_sans_flux}. Although it is $2\pi$-periodic, the plot between $-2\pi$ and $2\pi$ shows that, depending on the choice 
 of the minima modulo $2\pi$, the barrier between the equivalent minima can be different. Notice that if $g=1$, the $E_{TJ}(\varphi)$ curve
  possesses a cusp at $\varphi=\pi$, but this cusp is rounded by any small asymmetry between leads $2$ and $3$ ($g\neq1$).  

 In the case $g=1$, the energy minimum corresponding to the canted solution satisfies $\cos\frac{\varphi}{2}=\pm\frac{g_0}{2}$. It spans from 
 $\varphi=\pi$ for $g_0=0$, corresponding to a single $\pi$-junction $S_2-S_3$ through the dot, to $\varphi=\frac{2\pi}{3}$ or $\frac{4\pi}{3}$ 
 for $g_0=1$ (fully symmetric bijunction, Fig. \ref{fig:phase_vortices}a,b) and $\varphi=0$ for $g_0=2$ (Fig. \ref{fig:phase_vortices}c). If instead $g_0>2$, there is no canting
  and the bijunction displays two $\pi$-junctions $S_2-S_1$ and $S_1-S_3$ in series, and a $0$-junction $S_2-S_3$ (Fig. \ref{fig:phase_vortices}c).
 
As an essential fact, in the canted case there are two equivalent solutions, obtained by changing $\varphi$ in $-\varphi$ or in $2\pi-\varphi$ (Fig. \ref{fig:phase_vortices}a,b). As shown in Section III, the choice of the minima and of the corresponding barrier can be monitored by an external flux. 

This simple calculation shows that i) frustration manifests itself in canting the phases from $0$ or $\pi$; ii) a doubly degenerate state is formed, with opposite phase vorticities; iii) a too asymmetric 
bijunction does not sustain frustration, and yields two $\pi$-junctions and one $0$-junction. Phase vorticity appears as a spontaneous symmetry breaking, induced by the frustration brought by the existence of a localized spin creating $\pi$-junctions. The presence of the localized spin therefore induces a chirality in the bijunction. 

\subsection{A microscopic model}
Let us now provide a nonperturbative calculation, describing the localized spin with the help of a local Zeeman (or exchange field $J$), as in Ref. \onlinecite{Benjamin}. This can be related to a model including the Coulomb interaction through a mean-field approximation. This excludes the possible formation of a Kondo state in the $0$-junction regime when the dot-lead couplings are large enough. The Hamiltonian of the system is $H=H_{S}+H_{D}+H_{T}$ where $H_{S}$, $H_{D}$ and $H_{T}$ respectively denote the lead, dot and lead-dot tunneling contributions. The dot part is written as:

\begin{equation}
H_D=E_0\sum_{s=\uparrow,\downarrow} d^{\dagger}_s d_s - J(d^{\dagger}_{\uparrow} d_{\uparrow}-d^{\dagger}_{\downarrow} d_{\downarrow}).
\end{equation}
where $E_0$ is the bare energy level. We assume that $E_0-J<0$ and $E_0+J>0$, such that for weak coupling to the leads, the dot level carries one electron with spin up. Writing $H$ in the Nambu notation $H=H_{S}+H_{D}+H_{T}$, and performing a gauge transformation to incorporate the superconducting phases $\varphi_j$ in the tunneling term $H_T$, one gets, up to an additive constant, the following expressions:

\begin{equation}\label{Hamilt}
H_S=\sum_{j=1,2,3}\sum_{k}
\Psi_{jk}^{\dagger}(\xi_{k}\sigma_{z}+\Delta_j \sigma_{x})
\Psi_{jk}, \Psi_{jk}=\left(\begin{array}{c} \psi_{jk,
\uparrow}\\
\psi_{j(-k),\downarrow}^{\dagger}
\end{array} \right)
\end{equation}

\begin{equation}
H_{D}=d^{\dagger}(E_{0} \sigma_{z}-J \sigma_0)d
\label{eq:dspinor}
\end{equation}

\begin{equation}
 H_{T}=\sum_{jk}\Psi_{jk}^{\dagger} T_{j} d + h.c.,
\quad \quad  d=\left(\begin{array}{c} d_{\uparrow}\\
d_{\downarrow}^{\dagger}
\end{array} \right),
\label{eq:dspinor}
\end{equation}

\noindent 
with $ T_{j} =t_{j} \sigma_{z}e^{i\sigma_{z}\varphi_j/2}$ and $t_{j} $ is
the tunnelling amplitude between the lead $j$ and the dot. $\sigma_{0}$ is the identity matrix and $\sigma_{x,y,z}$ denote the Pauli matrices in the basis formed by electrons with spin $\uparrow$ and holes with spin $\downarrow$.

The procedure to obtain the Andreev bound states and the current-phase relationships by writing an effective action for the two dots can be found in Ref. [\onlinecite{Benjamin}]. One writes the partition function as 
 
\begin{align}
Z = \int \mathcal{D} \left[ \bar{\psi},\psi , \bar{d}, d\right] e^{- S \left[ \bar{\psi},\psi , \bar{d}, d \right]} ,
\end{align}
e.g. as a functional integral over Grassmann fields for
the electronic degrees of freedom ($\Psi_{jk}, \bar\Psi_{jk}, d, \bar d$). The
Euclidean action reads:

\begin{equation}
S_A=S_{D}+\int_{0}^{\beta} \! d \tau
[\sum_{jk}{\bar\Psi_{jk}(\tau)}(\partial_{\tau} \sigma_0+\xi_{k}\sigma_{z}+\Delta_j\sigma_{x})\Psi_{jk}(\tau)+{
H_{T}(\tau)}].
\end{equation}

\noindent $\beta$ is the inverse temperature, and
${H_{T}(\tau)}=\sum_{jk}{\bar\Psi_{jk}(\tau)}T_{j}d_{}(\tau)+h.c.$
while $S_{D}=\int_{0}^{\beta}\! d\tau [{\bar d(\tau)}(\partial_{\tau} \sigma_0+\epsilon_{}
\sigma_{z})d(\tau)]$.
 After integrating out the leads we get $Z=\int\! \mathcal{D} \left[\bar{d}, d\right] \; e^{-S_{eff}}$ 
 with
\begin{equation}
S_{eff}=S_{D}-\int_{0}^{\beta}\! d\tau \; d\tau' \;{\bar d(\tau)}{\check
\Sigma(\tau - \tau ')} d(\tau '),
\end{equation}
where $\check
\Sigma(\tau)=\sum_{j=1,2,3} T_{j}^{\dagger} G_{j}(\tau) T_{j}$ and
$G_{j}(\tau)=\sum_{k}
(\partial_{\tau} \sigma_0+\xi_{k}\sigma_{z}+\Delta_j\sigma_{x})^{-1}
\delta(\tau)$.

We perform a Fourier transform on the Matsubara frequencies (with  $\omega_{n}=(2n+1)\pi/\beta$):
$\delta(\tau)=\frac{1}{\beta} \sum_{\omega_n} e^{-i \omega_{n} \tau}$ and
$G(\tau)=\frac{1}{\beta} \sum_{\omega_n} e^{-i \omega_{n} \tau} G(i\omega_n)$, which gives for the Green function $G_j$:

\begin{eqnarray}
G_j(i\omega_n)=\int\! d \xi\; \nu(\xi)(-i
\omega_{n} \sigma_0+\xi_{k}\sigma_{z}+\Delta_j\sigma_{x})^{-1}
\\
\nonumber
\simeq \frac{\pi
\nu(0)}{\sqrt{\Delta_j^2-(i\omega_n)^2}}(i\omega_{n} \sigma_0+\Delta_j \sigma_{x})
\end{eqnarray}

\noindent 
Here $\nu(\xi)=\sum_{k}\delta(\xi-\xi_{k})$ is
approximated by a constant $\nu(0)$, the density of states at the
Fermi level in the normal leads. Let us assume for sake of simplicity the three gaps equal, $\Delta_j=\Delta$. One finally obtains the effective action
(introducing $d_{\alpha}(\tau)=\frac{1}{\sqrt{\beta}}\sum_{\omega_{n}} e^{-i\omega_{n} \tau} d_{\alpha}(i\omega_{n})$)

\begin{eqnarray}  \nonumber
S_{eff}=&\sum_{\omega_{n}} \bar{\bf d}(\omega_{n}){\bf {\cal M}}(i\omega_n){\bf d}(i\omega_{n}) \\
{\bf {\cal M}}(i\omega_n) = &(-i \omega_{n}+J) \sigma_{0} +E_0
\sigma_{z}-{\check{ \bf \Sigma}_{i\omega_{n}}},
\end{eqnarray}
${\bf {\cal M}}(i\omega_n)$ is described by a $2$ x $2$ matrix, whose coefficients are given by

\begin{widetext}
\begin{equation}\label{Matrix} 
\begin{aligned}
{\cal M}_{11}&=i\omega_n(1+\frac{\Gamma}{2\sqrt{\Delta^2-(i\omega_n)^2}})-E_0+J,\,\,\,\,\,\,
{\cal M}_{12}=-\frac{\Gamma \Delta}{2\sqrt{\Delta^2-(i\omega_n)^2}}(\sum_i\gamma_ie^{-i\varphi_i}),\\
{\cal M}_{21}&=-\frac{\Gamma \Delta}{2\sqrt{\Delta^2-(i\omega_n)^2}}(\sum_i\gamma_ie^{i\varphi_i}),\,\,\,\,\,\,
{\cal M}_{22}=i\omega_n(1+\frac{\Gamma}{2\sqrt{\Delta^2-(i\omega_n)^2}})+E_0+J,\\
\end{aligned}
\end{equation}
\end{widetext}

\noindent
with $\Gamma=2\pi \nu(0)\sum_i|t_i|^2$ and $\gamma_i=|t_i|^2/\sum_i|t_i|^2$. ${\bf {\cal M}}$ is an hermitian matrix once $i\omega_n$ is replaced by the real number $z$. The dispersion relation for the Andreev bound states is given by the eigenvalues of the effective action, replacing $i\omega_n$ by $z$. After integrating out the
$\{d,\bar d\}$ variables, the partition function is given by

\begin{equation}
Z=\int \! \mathcal{D} \left[ \bar{d}, d\right] \; e^{-S_{eff}}=\prod_{i\omega_n} \det
{\bf {\cal M}}(\omega_{n}).
\end{equation}

The free energy is given by:
\begin{equation}
 F=-\frac{1}{\beta}\sum_{\omega_n}\ln(\det {\bf {\cal M}}(i\omega_n)).
\end{equation}

The Josephson current in $S_i$ is expressed as:

\begin{align}
\nonumber
I_{Ji}=-\frac{2e}{\hbar\beta}\frac{\partial}{\partial \varphi_{i}}  \ln Z\\
=-\frac{2}{\beta}\frac{\partial}{\partial \varphi_{i}} \sum_{\omega_n} \ln
(\det {\bf {\cal M}}(i\omega_n))
\end{align}

\begin{figure}[htb]
\includegraphics[width=.6\columnwidth]{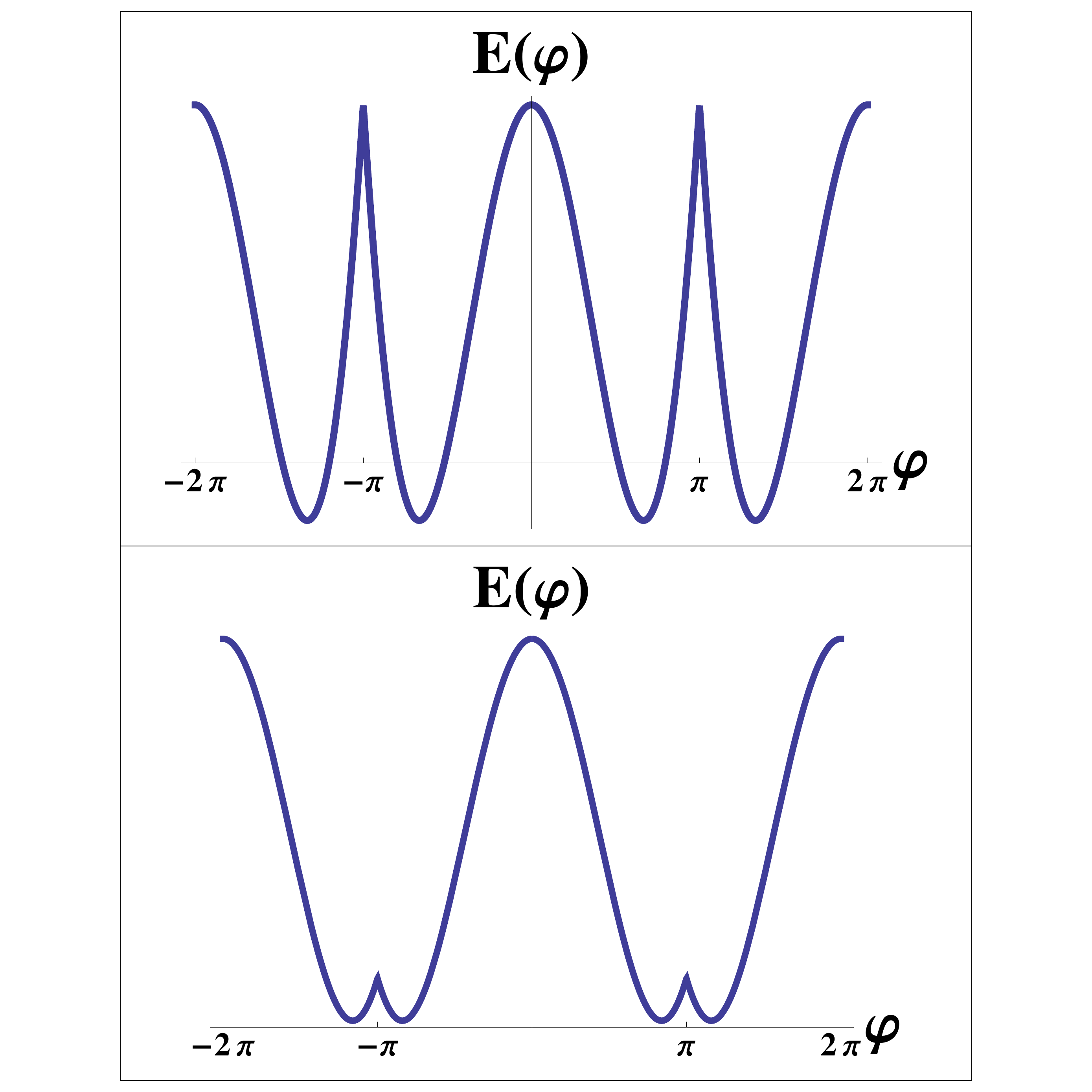}
\caption{(Color online). Total energy of the bijunction, from the microscopic model, as a function of the phase $\varphi_{23}=\varphi$. Parameters are $\Delta=1$, $\Gamma=2$, $J=5$, $\varepsilon=0$, temperature $T=0.02$, and (a) $\gamma_{1,2,3}=(1/3,1/3,1/3)$, (b) $\gamma_{1,2,3}=(0.4,0.4,0.2)$.
\label{fig:exact}
}
\end{figure}

Consider for simplicity the case of a bijunction symmetric by exchange of leads $2$ and $3$, e.g. $\gamma_2=\gamma_3$, to be compared 
with the $g=1$ case of Section I. If the exchange field is sufficient to stabilize a local moment, one also finds a critical value of the asymmetry 
$\frac{\gamma_1}{\gamma_{2,3}}$ above which frustration disappears and the bijunction is dominated by two $\pi$-junctions in series. In the
 perturbative limit where $\Gamma$ is smaller than the single spin level $|E_0-J|$, one finds energy profiles $E_{TJ}(\varphi)$ similar to those of the
  phenomenological model, with couplings $g_{ij}$ respectively proportionnal to $\gamma_i\gamma_j$. An example of an exact nonperturbative
   solution is given in Fig. \ref{fig:exact}. More generally, the critical value of $J$ above which the $\pi$-junctions are stabilized is about 
   $\frac{J}{\Gamma}=0.5$. In this regime, because the $\pi$-junction is weaker than a $0$-junction, the perturbative calculation turns out to 
   be qualitatively correct, and the physics is well described by the phenomenological model. 

\section{bijunction in a circuit with loops}
Superconducting interference devices with embedded junctions can be used to measure their phase-current relation \cite{Dam, Cleuziou, Rocca}. These techniques imply inserting the junction in a multiple connected circuit. The above analysis shows that the presence of three $\pi$-junctions can create frustration and phase canting at the junctions. Having three superconducting reservoirs, the bijunction can be connected in various ways to an external circuit. 

\subsection{Single loop}
\begin{figure}[htb]
\includegraphics[width=0.8\columnwidth]{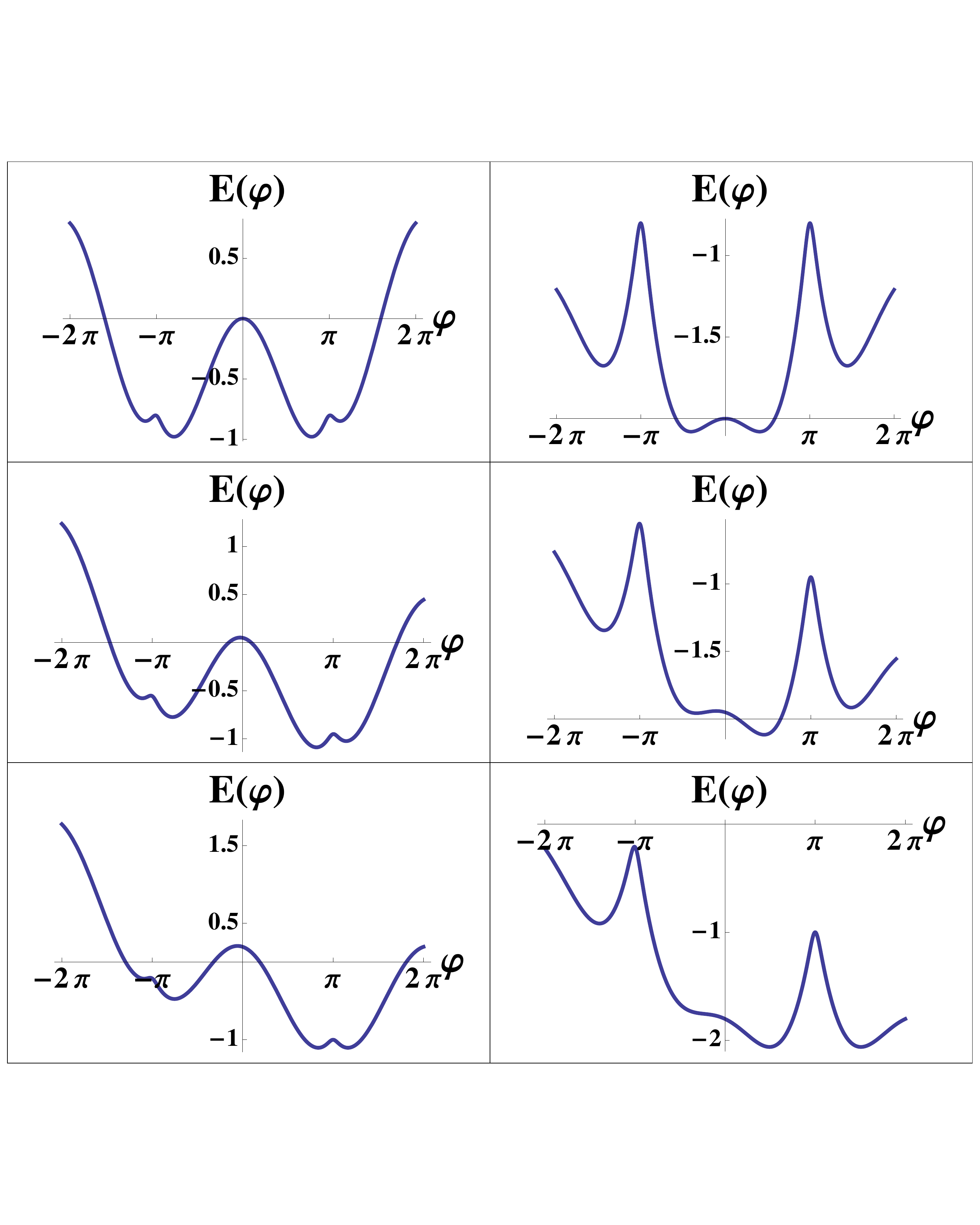}
\caption{(Color online). Total energy of the bijunction inserted in a single loop, as a function of the phase $\varphi_{23}=\varphi$. The asymmetry  betwen leads $2$ and $3$ is weak. Parameters are (left panels) $g_0=0.5$ and (right panels) $g_0=1.5$, and from top to bottom $\Phi_{ext}=0, \Phi_0/4, \Phi_0/2$.
\label{fig:phenomenol_avec_flux}
}
\end{figure}

Let us first consider here the simplest geometry obtained by connecting superconductors $S_2$ and $S_3$ by a loop, leaving superconductor $S_1$ disconnected. This implies that the phase $\varphi=\varphi_2-\varphi_3$ is accessible and controllable experimentally, while the "floating" phase $\varphi_1$ is determined by the condition $J_1=0$ (Equation \ref{zerocurrent}). Let us denote by $L$ the loop inductance, $\Phi_{ext}$ the external flux, and $LI$ the flux embedded in the loop, induced by the current $I$. Expressing flux quantification along the circuit enclosing the loop and passing through the dot yields :

\begin{equation}
\varphi=\frac{2\pi}{\Phi_0}\Phi [2\pi],
\end{equation}
where $\Phi=\Phi_{ext}+LI$ is the total embedded flux and $\Phi_0=hc/2e$ is the elementary flux quantum. The total energy becomes:

\begin{equation}
\label{eq:one loop}
E_{TJ1L}(\varphi)=\frac{\Phi_0^2}{8\pi^2L}(\varphi-\varphi_{ext})^2+E_{TJ}(\varphi)
\end{equation}
where $E_{TJ}$ is given by Equation \ref{eq:energy_TJ}, and with $\varphi_{ext}=\frac{2\pi}{\Phi_0}\Phi_{ext}$.
Consider first $\Phi_{ext}=0$. Then, if $L<L_c$, the only stable solution is $\Phi=0$, and there is no equilibrium phase difference at the
 junction. Conversely, if $L>L_c$, a spontaneous flux appears in the loop, together with a phase difference $\varphi'_m$ at the junction 
 (Fig. \ref{fig:phenomenol_avec_flux}). When $LI_c>>\phi_0$ ($I_c$ is the critical current of the bijunction), 
 $\Phi\simeq\pm\frac{\Phi_0}{2\pi}\varphi_m$ thus a large  loop stabilizes the two vortex solutions found in Section I. If on the contrary 
 $\Phi_{ext}=\Phi_0/2$, the loop stabilizes the solutions $\Phi\simeq\frac{\Phi_0}{2\pi}\varphi_m, \frac{\Phi_0}{2\pi}(2\pi-\varphi_m)$. These 
 two sets of solutions are equivalent, but the barrier between the two degenerate mimima are different. For a given set of the parameters 
 $g_0,g$, the highest barrier is encountered for one or the other of the applied fluxes. On the other hand, if $\Phi_{ext}$ is not a multiple of 
 $\Phi_0/2$, the minima are not equivalent (Fig. \ref{fig:phenomenol_avec_flux}). Fig. \ref{fig:phenomenol_avec_flux} shows that 
 one may keep the two minima at fixed values, say ($\pm \varphi_m$), and vary the asymmetry parameter $g_0$ 
 (Fig. \ref{fig:phenomenol_avec_flux}, top panels), thus changing the barrier between the two minima. More interestingly, one can keep the same 
 bijunction parameters and change the flux from $\Phi_{ext}=0$ to $\Phi_{ext}=\Phi_0/2$ (Fig. \ref{fig:phenomenol_avec_flux}, left panels, 
 or right panels). This switches the pair of minima from  ($\pm \varphi_m$) to ($\varphi_m$, $2\pi-\varphi_m$), with a strong change of the 
 barrier between them. Depending on whether $g_0$ is smaller or larger than $1$, the barrier may be decreased or increased.

This might offer a way of manipulating the pair of vortex solutions as a phase qubit, by tuning from three to two energy minima. Actually, tuning the flux between $\Phi_{ext}=0$ and $\Phi_{ext}=\Phi_0/2$ keeps two of the three states equally probable but allow to switch on or off the tunneling between them. On the other hand, fixing the flux to a value such as  $\Phi_{ext}=\pm\Phi_0/4$ favours one or the other minima. 

The above discussion shows that for a moderate asymmetry and a large inductance, this set-up allows a spontaneous current/flux to appear in the loop. Contrarily to the simple $\pi-$ junction where only a flux $\Phi_0/2$ can be stabilized, here the induced flux can take any value between $0$ and $\Phi_0$. 

\begin{figure}[htb]
\includegraphics[width=1.0\columnwidth]{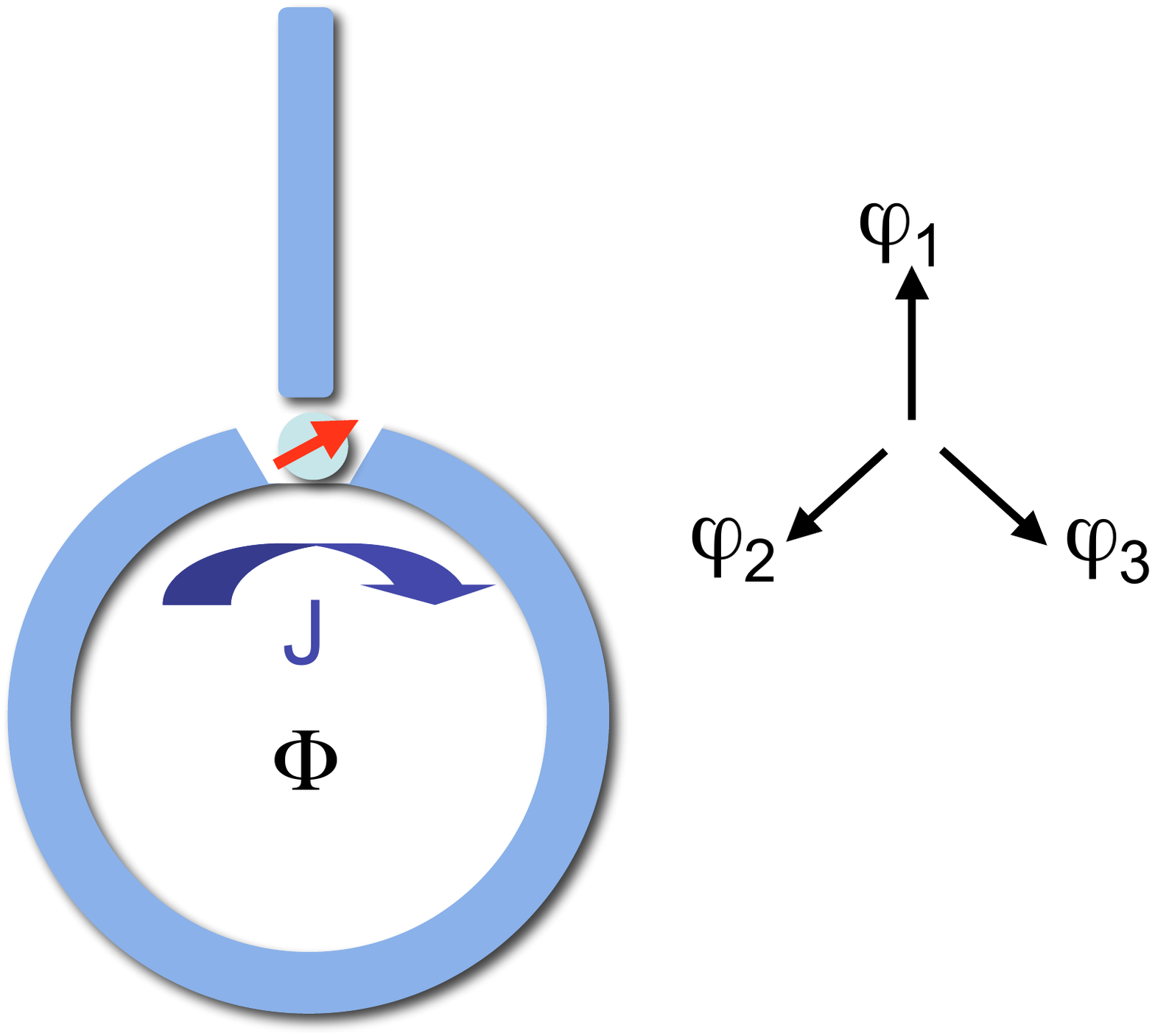}
\caption{(Color online).Connecting the bijunction with one loop can stabilize an arbitrary flux. $J$ (blue arrow) 
denotes the current flowing in junction $S_2-S_3$.
\label{fig:one_loop}
}
\end{figure}

\subsection{Two loops}
\begin{figure}[htb]
\includegraphics[width=1.0\columnwidth]{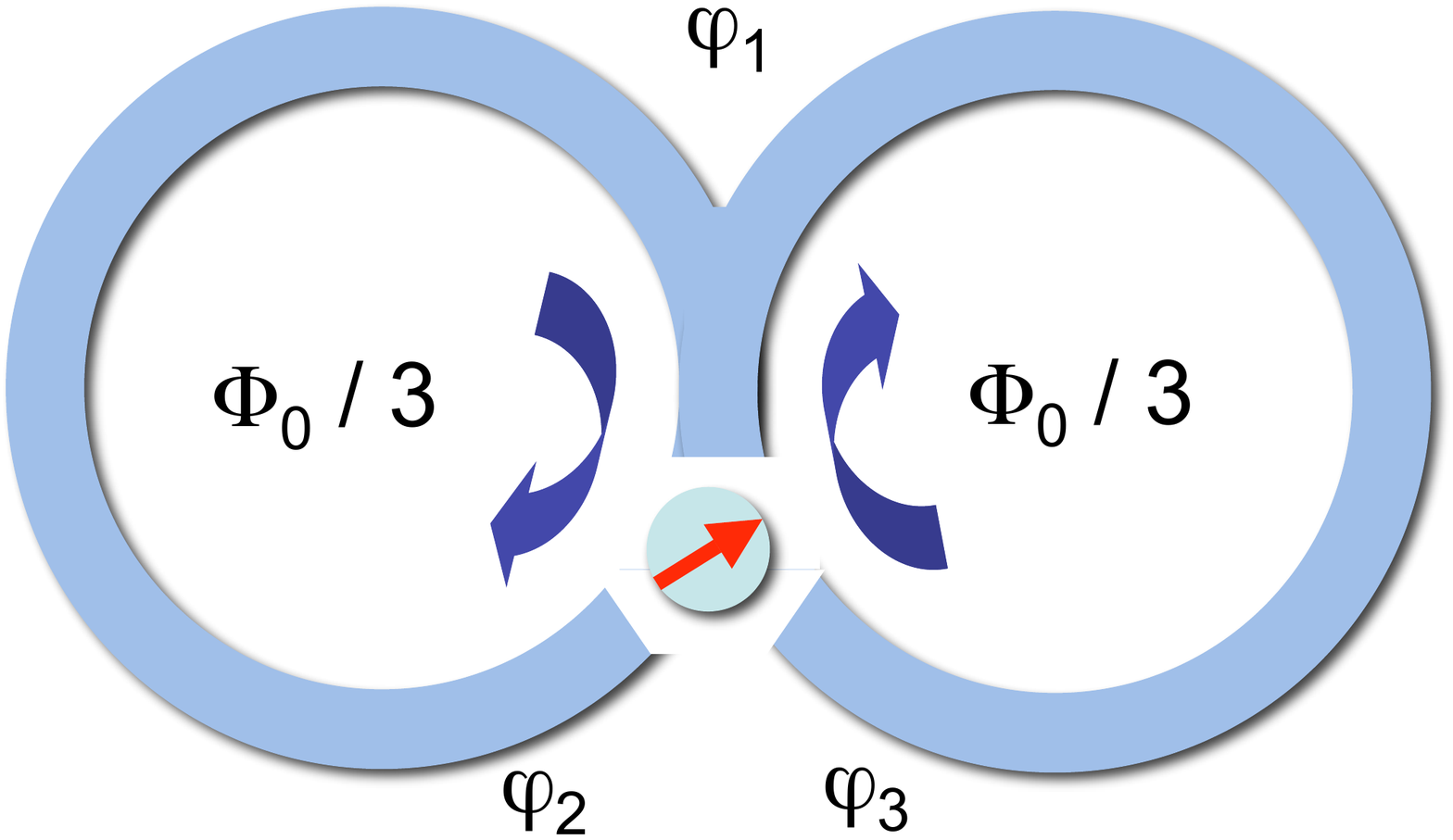}
\caption{(Color online).Connecting the bijunction with two loops stabilizes two symmetric spontaneous fluxes. The Figure corresponds to zero external flux, large inductance and symmetric bijunction. The blue arrows denote the currents circulating in the junctions.
\label{fig:two_loop}
}
\end{figure}

Let us now connect the bijunction by two loops, by closing for instance  the junctions $S_1-S_2$ and $S_1-S_3$ (Fig. \ref{fig:two_loop}). Those loops respectively enclose fluxes $\Phi_{ext,3}$ and $\Phi_{ext,2}$. The quantification condition for each of the loops are:

\begin{equation}
\varphi_{12}=\frac{2\pi}{\Phi_0}\Phi_{3} [2\pi], \;\;\varphi_{13}=-\frac{2\pi}{\Phi_0}\Phi_{2} [2\pi],
\end{equation}
where $\Phi_{2}=\Phi_{ext,2}+LI_{13}$ and $\Phi_{3}=\Phi_{ext,3}-LI_{12}$ are the total embedded flux. Defining $\varphi_{ext,2}=\frac{2\pi}{\Phi_0}\Phi_{ext,2}$, $\varphi_{ext,3}=\frac{2\pi}{\Phi_0}\Phi_{ext,3}$, the total energy then reads :

\begin{equation}
\begin{aligned}
E_{TJ2L}(\varphi)=\frac{\Phi_0^2}{8\pi^2L}\Big[(\varphi_{12}-\varphi_{ext,3})^2
+(\varphi_{13}+\varphi_{ext,2})^2\Big]\\
+E_{0}[g_0g\cos\varphi_{12}+g_0\cos\varphi_{13}+\cos(\varphi_{13}-\varphi_{12})].
\end{aligned}
\end{equation}

If $L$ is large, minimizing with respect to $\varphi_{12}$, $\varphi_{13}$ gives in the frustrated regime the two symmetric vortex 
solutions of Section I, which induce nonzero but equal fluxes in the loops. The fluxes can take the values 
$\Phi_2=\Phi_3\simeq\frac{\Phi_0}{2\pi}\varphi_m$,  or $\Phi_2=\Phi_3\simeq-\frac{\Phi_0}{2\pi}\varphi_m$ (Fig. \ref{fig:two_loop}). 
In the symmetric junction case, fluxes $(\Phi_0/3,\Phi_0/3)$ or $(-\Phi_0/3,-\Phi_0/3)$ can be stabilized. Those flux can be made 
dissymetric either by acting on the junction parameters (with gates) of with en external flux.

\subsection{Three loops}
\begin{figure}[htb]
\includegraphics[width=1.0\columnwidth]{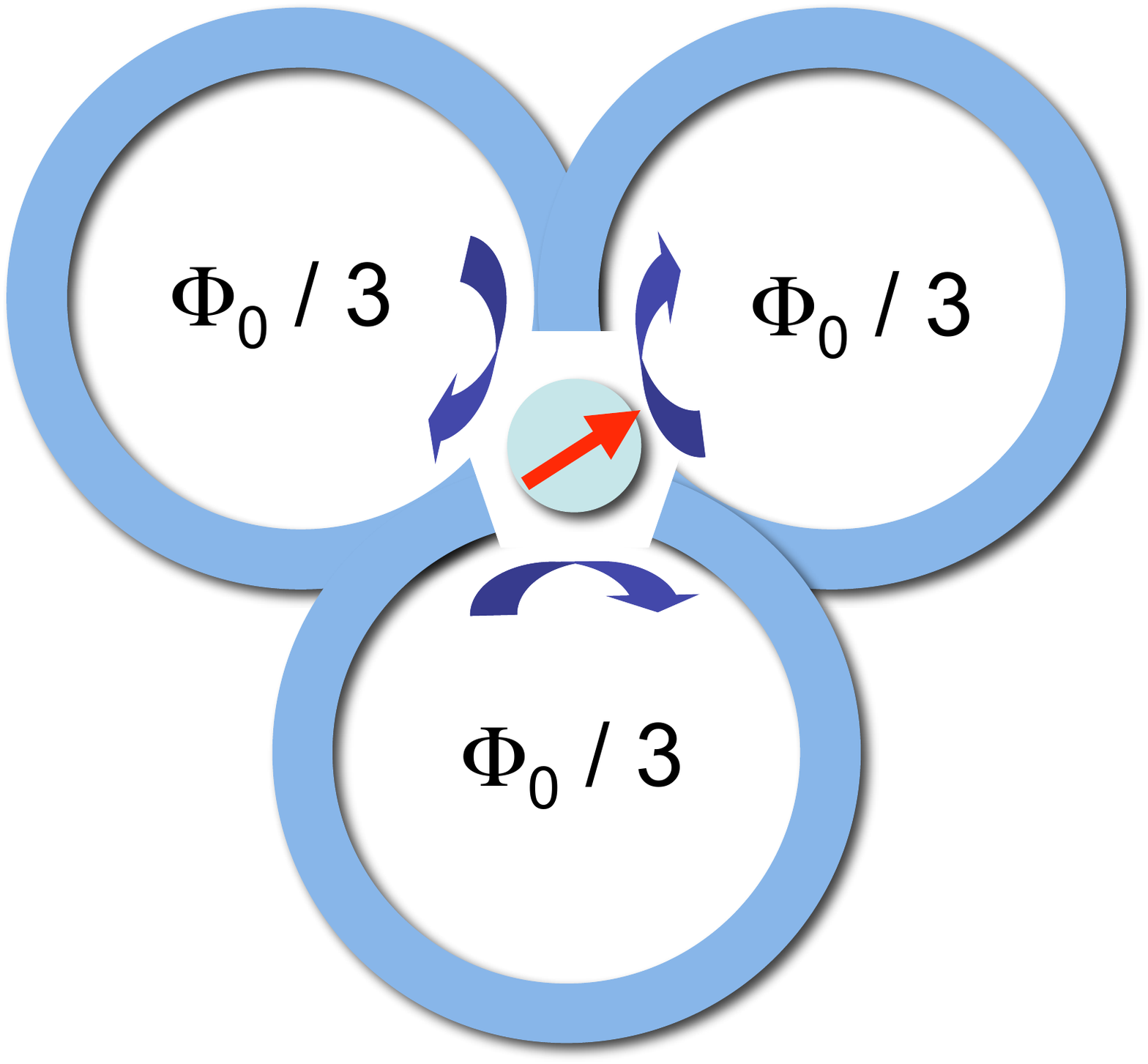}
\caption{(Color online).Connecting the bijunction with three loops globally traps one flux quantum.
\label{fig:three_loops}
}
\end{figure}

Finally, the bijunction can be more symmetrically closed by three loops, each embedding an external flux $\Phi_{ext,i}$ (Fig. \ref{fig:three_loops}). Thus ($i=1,2,3)$:

\begin{equation}
\varphi_{ij}=\frac{2\pi}{\Phi_0}\Phi_{k}\delta_{ijk} [2\pi], \;\;\Phi_{k}=\Phi_{ext,k}+L_{ij}\delta_{ijk},
\end{equation}
where $\delta_{ijk}$ is $1$ if all i,j,k are different, and zero otherwise.
The total energy reads :

\begin{equation}
\begin{aligned}
E_{TJ3L}(\varphi)=\frac{\Phi_0^2}{8\pi^2L}\sum_{ij}(\varphi_{ij}-\varphi_{ext,k}\delta_{ijk})^2\\
+E_{0}[g_0g\cos\varphi_{12}+g_0\cos\varphi_{13}+\cos(\varphi_{13}-\varphi_{12})].
\end{aligned}
\end{equation}

Large $L$ either yields $\Phi_1=\Phi_2=\Phi_3\simeq\frac{\Phi_0}{2\pi}\varphi_m$,  or $\Phi_1=\Phi_2=\Phi_3\simeq-\frac{\Phi_0}{2\pi}\varphi_m$. For instance, in the fully symmetric case, each loop carries one third of the flux quantum $\Phi_0$. A property of the three-loop configuration is that it globally embeds one flux quantum, $\Phi_1+\Phi_2+\Phi_3=\pm\Phi_0$. This is a direct manifestation of the phase vorticity induced by frustration. 

\section{conclusion}
We have shown that a quantum dot carrying a $1/2$ spin, thus able to generate a Josephson $\pi$ junction, may induce frustration if inserted in a Josephson bijunction. If the coupling between two superconductors - say $S_2$ and $S_3$ - dominates, this results in the $S_2-S_3$ junction being a $\pi$-junction, while low asymmetry leads to frustration and canting of the equilibrium phases. On the opposite, a too small coupling between $S_2$ and $S_3$ results in both $S_1-S_2$ and $S_1-S_3$ being $\pi$-junctions, making $S_2-S_3$ an effective $0$-junction.  The possibility of continuously tuning the junction $S_2-S_3$ between a $0$- and a $\pi$-junction is a first result of this work. 

This phenomenon displays an interesting link between two kins of magnetism : spin magnetism in the dot, and orbital magnetism manifesting in spontaneous flux (vortex). It is remarkable that this is a topological property, related to the existence of a localized spin, and not to its direction. In fact the direction of the localized spin and the sign of the stabilized vortex are unrelated.  This could be different in more complicated situations involving an additional spin-orbit coupling. 

The second result is that frustration generates two equivalent states possessing opposite phase vorticities, each of them breaking time-reversal symmetry. These states can be revealed by inserting the bijunction in a set-up containing one, two or three loops. In the case of a single loop, the two phase vortex states result in a spontanous flux crossing the loop, which is different from $0$ or $\pi$. While a zero external flux, or a multiple of $\Phi_0/2$, preserve the symmetry of the two vortex states, any other value lifts the degeneracy and can be used to stabilize one or the other of these two states. 

This might have some consequences in terms of using the above device for generating flux qubits ot flux qutrits \cite{Clarke,Langford,Buisson}. Indeed, in the two-loop scheme one may control the two distinct phases by the external fluxes and the bijunction parameters. Tunneling through the barrier separating the two states can be strongly varied if acting on, say, the coupling between lead $1$ and the dot, by split gates for instance. Quantum fluctuations of the trapped fluxes occur if the lead-dot junctions have finite capacitances, for instance if the reservoirs are Cooper pair boxes. Control of the "longitudinal" and "transversal" components of this flux qubit is thus possible, as a basic ingredient for applications. Further investigations must be carried out to derive an effective qubit model and check its feasibility.

The authors acknowledge the support of PICS CNRS-CONICET 5755, PICT 2010-1060 from ANPCyT, PIP 11220080101821 from CONICET, Argentina, grant 06-C400 from UNCuyo and the Laboratoire Franco-Argentin en Nanosciences (LIFAN).

\end{document}